\documentclass[12pt,a4 paper, one side]{article}
\usepackage{amsmath,amssymb,latexsym}
\usepackage{hyperref}
\usepackage{graphicx}
\usepackage{epstopdf}
\begin{document}
\date{}
\title{Dynamical model of expansion free dissipative perfect fluids in general relativity}
\author{Rajesh Kumar\footnote{rkmath2009@gmail.com}\footnote{rajeshkumar.mathstat@ddugu.ac.in} and S. K. Srivastava\footnote{sudhirpr66@rediffmail.com}\\
Department of Mathematics $\&$ Statistics,\\
Deen Dayal Upadhyaya Gorakhpur University, Gorakhpur, INDIA.}
\maketitle
\begin{abstract}
This paper  deals with the spherically symmetric self-gravitating star which is considered to be expansion free dissipative perfect fluids distribution. Some recent research reveals that expansion free dynamical star must be accelerating and dissipating. We adopted  some conjectures  to obtain the analytical solution for the dynamical model of such stars. Firstly, it has shown that density of dynamical star is homogeneous and $\Lambda-$ dominated under quasi-static diffusion approximation.  Secondly, the self-similar solution is also discussed to describe the dynamical model.
\end{abstract}
\section{Introduction}
The study of  evolution of self-gravitating celestial objects (e.g. star, galaxies, etc.) are the most fascinating  among relativists and astrophysicists. During the processes, self-gravitating objects may pass through phases of intense dynamical activities which can be observed by exploring the dynamical equations\cite{kj13}, \cite{sk09}. During the evolution of compact stars after explosion under the expansion-free condition, an interesting phenomenon of cavity formation has been observed. The evolution of expansion free self-gravitating object provides interesting cosmological significances and it consents to for the obtention of a wide range of solutions in general relativity and cosmology (\cite{hs97} - \cite{sy12a} and references their in).

\par
In 1960, Skripkin \cite{s1960} studied the very fascinating problem of the evolution of a spherically symmetric fluid distribution following a central explosion which result a Minkowskian cavity  surrounding the centre of fluid distribution. Herrera et al. \cite{hds08} discussed this problem by showing that under Skripkin conditions (constant energy density) the scalar expansion vanishes. It was further shown that the assumption of vanishing expansion scalar requires the existence of a vacuum cavity within the fluid distribution (of any kind).   A systematic study on shearing expansion-free spherically symmetric distributions was presented in \cite{hgs09} and  shown that in general for any non-dissipative fluid distribution, the expansion-free condition requires the energy density to be inhomogeneous. Recently, some authors ( \cite{phosc11}, \cite{ks2018}) have obtained interesting results for the spherically-symmetric self gravitating expansion-free   fluids with a vacuum-cavity. Some works (\cite{hpo10}, \cite{phosc11} - \cite{sa14}) were presented the dynamical stability of the expansion-free spherical collapse. Sharif and Yousaf have studied  the inhomogeneous non-dissipative dust model  with expansion free conditions and the model is completely integrable \cite{sy12}. Some analytical solutions for the self gravitating spherically-symmetric dissipation-less fluids with the expansion free motion have been obtained Di Prisco et al. \cite{phosc11}. Recently, Kumar and Srivastava(\cite{ks2018}, \cite{ks2018a}) have studied the gravitational collapse of self-gravitating star (dissipative/non-dissipative) which discussed  the model of vacuum cavity, which showed the potential of expansion free motion.  Recently, Sherif et al. \cite{sgm19} have studied the general features of  expansion free dynamical star and showed that  non-zero acceleration and dissipation are necessary for the evolution of such kinds of star. In present studies, authors are interested in the dynamics of the self-gravitating radiating (dissipative perfect fluid) star only in its evolution once cavity is already formed and The main contribution of this work to investigate  the dynamical solution of such relativistic expansion free star.

%%%%%%%%%%%%%%%%%%%%%%%%%%%%%%%%%%%%%%%%%%%%%%%%%%%%%%%%%%%%%%%%%%%%%%%%%%%%%%%%%%%%%%
\section{Fluids distribution, Einstein's field equations, Kinematical parameters and the junction conditions}
Consider a spherically symmetric radiating compact star, which is closed by spherically hypersurface $\Sigma^e$. The matter distribution is taken to be perfect fluid concerning dissipation in form of heat flow (diffusion approximation) and outgoing null fluid (streaming out limit). For the bounded or closed system, a space time line-element concerning to the fluids inside hypersurface $\Sigma^e$,
\begin{equation}
ds_{-}^2 = -A(t,r)^2dt^2 + B(t,r)^2 dr^2 + R(t,r)^2 (d\theta^2 + \sin\theta^2 d\phi^2)
\label{eq1}
\end{equation}
where the metric coefficient $B$ is dimension-less and $R$ defines the areal-radius of spherically-symmetric surface and $\dot{R}<0$ in case collapsing configuration. The coordinates is label as $x^\lambda = (t,r,\theta,\phi)$, $\lambda =0,1,2,3$. 
\par
Here the fluids - distribution are considered to be dissipative  perfect fluid expressed by the energy-momentum tensor (inside $\Sigma^e$ if the system being  closed/bounded)
\begin{equation}
T^{\alpha}_{\beta} = (p+\rho) v^\alpha v_\beta + p \delta^\alpha_\beta+q^\alpha v_\beta+v^\alpha q_\beta + \epsilon l^\alpha l_\beta
\label{eq2}
\end{equation}
where $p \rightarrow$ pressure, $\rho \rightarrow$ energy-density and $\epsilon \rightarrow$  radiation energy. And  $q_\alpha \rightarrow$  heat-flux,  $v_\alpha \rightarrow$  four-velocity vector  and $l_\alpha \rightarrow$ is a radial null four vector  which   satisfies
\begin{equation}
v^\alpha v_\alpha = -1, \quad~ v^\alpha q_\alpha = 0, \quad~ l^\alpha v_\alpha = -1,\quad~ l^\alpha l_\alpha = 0
\label{eq3}
\end{equation}
and system of  the comoving coordinate,
\begin{equation}
v^\alpha = A^{-1} \delta^\alpha_0, \quad~ q^\alpha = q B^{-1} \delta^\alpha_1, \quad~ l^\alpha = A^{-1} \delta^\alpha_0 +B^{-1} \delta^\alpha_1
\label{eq4}
\end{equation}
where $q$ is a function of $t$ an $r$ and $q^\alpha =  q \chi^\alpha$, $\chi^\alpha$ is a unit four vector along radial direction, satisfying
\begin{equation}
\chi^\alpha \chi_\alpha = 1, \quad~ \chi^\alpha v_\alpha = 0, \quad~ \chi^\alpha = \frac{1}{B} \delta^\alpha_1
\label{eq5}
\end{equation}
%%%%%%%%%%%%%%%%%%%%%%%%%%%%%%%%%%%%%%%%%%%%%%%%%%%%%%%%%%%%%%%%%%%%%%%%%%%%%%%%%%%%%%%%%%%
\begin{center}
\textbf{I. Einstein field equations}
\end{center}
Consider the evolution of cavity inside spherically symmetric self-gravitating radiating star. The general formalism of such dynamics is deployed details in  \cite{hds08}, \cite{ks2018}- \cite{ks2018a}. Following the work  it can be seen that as the consequence of the vanishing expansion scalar,   the line element (\ref{eq1}) reduces to the form
\begin{equation}
ds_{-}^2 = -A^2 dt^2 + R^{-4} dr^2 + R^2 (d\theta^2 +\sin\theta^2 d\phi^2)
\label{eq6}
\end{equation}
Thus, the Einstein's field equations  
\begin{equation*}
R^i_j - \frac{1}{2} \Re \delta^i_j = 8\pi T^i_j
\end{equation*}
yield following,
\begin{equation}
8\pi \tilde{\rho} A^2 = -3\frac{\dot{R}^2}{R^2}-A^2 R^4 (-\frac{1}{R^6}+5\frac{R'^2}{R^2}+2\frac{R''}{R})
\label{eq7}
\end{equation}
\begin{equation}
8\pi \tilde{q} \frac{A}{R^2}= 2(-\frac{A' \dot{R}}{A R}-2\frac{R' \dot{R}}{R^2}+\frac{\dot{R'}}{R})
\label{eq8}
\end{equation}
\begin{equation}
8\pi \tilde{p}  A^2 = -\frac{A^2}{R^2}+A^2 R' R^3 (2\frac{A'}{A}+\frac{R'}{R})-2\frac{\ddot{R}}{R}+\frac{\dot{R}}{R}(2\frac{\dot{A}}{A}-\frac{\dot{R}}{R})
\label{eq9}
\end{equation}
\begin{equation}
8 \pi p A^2 = A^2 R^4 [ 2\frac{A' R'}{A R} + \frac{R'}{R} (\frac{A'}{A}+2\frac{R'}{R})+\frac{A''}{A}+ \frac{R''}{R} ] -\frac{\dot{A} \dot{R}}{A R} -4 \frac{\dot{R}^2}{R^2} + \frac{\ddot{R}}{R}
\label{eq10}
\end{equation}
where $\tilde{\rho} = \rho+\epsilon$, $\tilde{p} = p+\epsilon$ and $\tilde{q} = q+\epsilon$. The Bianchi identities $T^{\alpha\beta}_{;\beta} = 0$ for the space time metric (\ref{eq6}),
\begin{equation}
\dot{\tilde{\rho}} - 2 \epsilon \frac{\dot{R}}{R} + \tilde{q}' R^2+ 2 R^2 \tilde{q} \frac{(AR)'}{AR} = 0
\label{eq11}
\end{equation}
\begin{equation}
\frac{\dot{\tilde{q}}}{A} - 2 \frac{\tilde{q}}{A} \frac{\dot{R}}{R} + R^2 [ \tilde{p}'+\frac{A'}{A} (p+\rho+2\epsilon) + 2 \epsilon \frac{R'}{R} ] = 0
\label{eq12}
\end{equation}
%%%%%%%%%%%%%%%%%%%%%%%%%%%%%%%%%%%%%%%%%%%%%%%%%%%%%%%%%%%%%%%%%%%%%%%%%%%%%%%%%%%%%%%%%%%%%%%
\begin{center}
\textbf{II. Mass-function, Kinematic-parameters and  Weyl curvature tensor}
\end{center}
The mass-function $m(t,r)$ introduced in ~\cite{ms64}, is  for space time metric (\ref{eq6}) is 
\begin{equation}
m(t,r) = \frac{1}{2} R (\frac{\dot{R}^2}{A^2} - R^4 R'^2 +1)
\label{eq13}
\end{equation}
In view of the equations (\ref{eq7}) - (\ref{eq9}), one can obtain from (\ref{eq13})
\begin{equation}
\dot{m} = - 4\pi ( \tilde{p} \dot{R} + \tilde{q} A R' R^2 )
\label{eq14}
\end{equation}
\begin{equation}
m' = 4\pi  ( \tilde{\rho} R' R^2 + \tilde{q} \frac{\dot{R}}{A} )
\label{eq15}
\end{equation}
On integration of Equ(\ref{eq15}) gives
\begin{equation}
3\frac{m}{R^3} = 4\pi \tilde{\rho} -4\pi \int R^3 (\tilde{\rho'}-3\frac{\tilde{q}\dot{R}}{A R^3}) dr 
\label{eq15a}
\end{equation}
The acceleration vector ($a^\alpha$)  and shear tensor ($\sigma_{\alpha\beta}$)  are defined by 
\begin{equation}
a^\alpha = v^\alpha_{;\beta} v^\beta
\label{eq15b}
\end{equation}
\begin{equation}
\sigma_{\alpha\beta} = \frac{1}{2} (v_{\alpha;\beta} + v_{\beta;\alpha})
\label{eq15c}
\end{equation}
For the metric (\ref{eq6}) above yields,
\begin{equation}
a^\alpha = a \chi^\alpha = (0, a R^2, 0,0), \quad~ a = \frac{A'}{A} R^2
\label{eq16}
\end{equation}
and,
\begin{equation}
  \sigma = -3\frac{\dot{R}}{ AR}
\label{eq17}
\end{equation}
where $\sigma^2 = \frac{3}{2} \sigma^{\alpha\beta} \sigma_{\alpha\beta}$.Also, it can be find from Equ (\ref{eq8}) and (\ref{eq17}),
\begin{equation}
4\pi \tilde{q}+\frac{1}{3} \sigma' R^2 +\sigma R R' = 0
\label{eq18}
\end{equation}
The Weyl curvature tensor ($C^\rho_{\alpha\beta\gamma}$) is defined via the Riemann curvature tensor $R^\rho_{\alpha\beta\gamma}$, the Ricci tensor $R_{\alpha\beta}$ and  scalar curvature  $\Re$ in four-dim space-time
\begin{equation}
C^\rho_{\alpha\beta\gamma} = R^\rho_{\alpha\beta\gamma} -\frac{1}{2}(R^\rho_\beta g_{\alpha\gamma} -R_{\alpha\beta} \delta^\rho_\gamma + R_{\alpha\gamma} \delta^\rho_\beta -R^\rho_\gamma g_{\alpha\beta}) + \frac{1}{6} \Re (\delta^\rho_\beta g_{\alpha\gamma} - g_{\alpha\beta} \delta^\rho_\gamma)
\label{eq19}
\end{equation}
The electric-Weyl tensor (magnetic-Weyl tensor absent due to spherically-symmetry of the  metric) is given by
\begin{equation}
E_{\alpha\beta} = C_{\alpha\gamma\beta\nu} v^\gamma v^\nu
\label{eq20}
\end{equation}
It can be seen that the electric Weyl tensor $E_{\alpha\beta}$ can also be expressed as \cite{hpoc10}
\begin{equation}
E_{\alpha\beta} = E(\chi_\alpha \chi_\beta -\frac{1}{3} h_{\alpha\beta})
\label{eq21}
\end{equation}
 where  $h^\alpha_\beta = g^\alpha_\beta + v^\alpha v_\beta$
\begin{equation}
2 E = -\frac{1}{R^2} + R^4 [ \frac{A''}{A} -\frac{R''}{R}+\frac{R'}{R}(\frac{A'}{A}-\frac{R'}{R}) ]+ \frac{1}{A^2} [ 3\frac{\ddot{R}}{R}-9\frac{\dot{R}^2}{R^2}-3\frac{\dot{R}}{R} \frac{\dot{A}}{A} ]
\label{eq22} 
\end{equation}
Also, in view of Equ. (\ref{eq7}), (\ref{eq13}) and (\ref{eq22}) it yield \cite{hpoc10}
\begin{equation}
3\frac{m}{R^3} = 4\pi \rho- E
\label{eq23}
\end{equation}
Sherif et al.\cite{sgm19} prove that expansion free dynamical star must be conformally flat $E=0$, then we have
\begin{equation}
8 \pi \rho = \frac{6m}{R^3}
\label{eq23a}
\end{equation}

%%%%%%%%%%%%%%%%%%%%%%%%%%%%%%%%%%%%%%%%%%%%%%%%%%%%%%%%%%%%%%%%%%%%%%%%%%%%%%%%%%%%%%%%%%%%
\begin{center}
\textbf{III. The Junction conditions}
\end{center}
For the bounded configuration (self-gravitating system), exterior of   spherically-symmetric surface $\Sigma^e$ (on $r=r_{\Sigma^e}$), the Vaidya  metric ( it has been assumed that all the outgoing-radiation is massless ) is given by
\begin{equation}
ds^2 = -(1-\frac{2M}{\rho}) d\tau^2 -2d\rho d\tau+ \rho^2 (d\theta^2+\sin\theta^2 d\phi^2)
\label{eq26}
\end{equation}
$M = M(\tau)$ denote the  mass , and $\tau$  the retarded-time.
\par
Since the expansion free models construct an central vacuum-cavity (Minkowski space time) of the fluids distribution. If  $\Sigma^i$ (on $r=r_{\Sigma^i}$) denotes  boundary-surface between the vacuum-cavity and the fluids, then such evolution problem requires Darmois-junction conditions  connect two distinct space-time metric into one such that the  both the hypersurfaces $\Sigma^e$ and $\Sigma^i$   demarcate the  distribution of fluids. The former causes distribution of fluids from Vaidya metric and the later distinguish from  space-time metric of vacuum cavity.
\par
The junction of  space-time (\ref{eq6}) to the Vaidya-metric (\ref{eq26}) and Minkowskian space-time over the  surface $\Sigma^e$ and $\Sigma^i$ respectively  has  studied in  \cite{hds08}- \cite{hpo10},\cite{ ks2018},  \cite{C00} and \cite{S85} ,  as

\begin{equation}
m(t, r_{\Sigma^e}) =  M(\tau), \quad~ q(t, r_{\Sigma^e}) =  p(t,r_{\Sigma^e})
\label{eq27}
\end{equation}
\begin{equation}
m(t, r_{\Sigma^i}) =  0, \quad~ q(t, r_{\Sigma^i}) = p(t, r_{\Sigma^i})
\label{eq28}
\end{equation}
It is also interesting to note that during the expansion free evolution, the matter of the voids stream out, which decreases the density of void from inside and becomes zero at $\Sigma^i$ (it follows from (\ref{eq23a}), $\rho(t,r_{\Sigma^i}) = 0$). If it allows for the presence of thin shell on $\Sigma^i$, then it has to relax the junction condition (\ref{eq28}) and consent to the discontinuity of the mass-function (\cite{phosc11}).
%%%%%%%%%%%%%%%%%%%%%%%%%%%%%%%%%%%%%%%%%%%%%%%%%%%%%%%%%%%%%%%%%%%%%%%%%%%%%%%%%%%%%%%%%%%%%%%%%%%%%%%%%
%%%%%%%%%%%%%%%%%%%%%%%%%%%%%%%%%%%%%%%%%%%%%%%%%%%%%%%%%%%%%%%%%%%%%%%%%%%%%%%%%%
\section{Diffusion approximation}
A renewed attention in general relativity is that self gravitating gravitational collapsing system is a dissipative-process. Dissipation-processes are usually considered with two feasible approximations: diffusion-approximation and streaming-out limit (\cite{is76} - \cite{l88}). The diffusion-approximation is rational, since it pertains whenever the mean-free path of fluids responsible for the energy-propagation is little as compare with the usual length of the system, which instituted very regular in the cosmological scenarios.
\par
This section  discussed the  purely diffusion approximation case which reveal  that $q \neq 0, \epsilon = 0$  considering with quasi-static regime\cite{hpoc10}. The quasi static means that the sphere changes slowly in time scale that is very long compared to the typical time in which the sphere reacts to a straight perturbation of hydrostatic equilibrium and so in this case  $\ddot{A} = \ddot{R}=\dot{A} \dot{R} = \dot{R}^2= 0 $~\cite{hbps02}. Then from equ. (\ref{eq9}) -  (\ref{eq10})

\begin{equation}
R^4 (\frac{A'R'}{AR} + \frac{R'^2}{R^2} +\frac{A''}{A}+\frac{R''}{R})+\frac{1}{R^2} = 0
\label{d1}
\end{equation}
Taking use of (\ref{eq23a}), it gives
\begin{equation}
\frac{A''}{A} + \frac{A' R'}{A R} = 0
\label{d2}
\end{equation}
which gives after integration 
\begin{equation}
A = \int \frac{d_1(t)}{R} dr + d_2(t)
\label{d3}
\end{equation}
where $d_1$ and $d_2$ are arbitrary function of $t$. Introducing \ref{d3} into  (\ref{d1}) gives
\begin{equation}
R^5 R'' + R^4 R'^2 +1 =0
\label{d4}
\end{equation} 
The first integral of (\ref{d4}) gives
\begin{equation}
R' = \frac{\sqrt{1+d_3(t) R^2}}{R^2}
\label{d5}
\end{equation}
where $d_3(t)$ is an arbitrary function. Further, integration of (\ref{d5}) yields
\begin{equation}
R \frac{\sqrt{1+d_3 R^2}}{2d_3} -  \frac{Sinh^{-1}(R\sqrt{d_3})}{2 d_3^{\frac{3}{2}}} = d_4 + r
\label{d6}
\end{equation}
which determine $R(t,r)$,where $d_4$ is an integrating constant and $d_3  >0$. Thus, in view of equs. (\ref{d4}) - (\ref{d6}),  equations (\ref{eq7}) - (\ref{eq9}) yield
\begin{equation}
8 \pi \rho = -3 d_3 (t) = \rho(t)
\label{d7}
\end{equation}
\begin{equation}
8 \pi p = 2 d_1 \frac{\sqrt{1+d_3 R^2}}{\int \frac{d_1}{R} dr + d_2} + d_3
\label{d8}
\end{equation}
\begin{equation}
8 \pi q = \frac{-2 \dot{R}}{\int \frac{d_1}{R} dr +d_2} [\frac{d_1}{\int \frac{d_1}{R} dr +d_2}+\frac{(4+3d_3 R^2)}{R^2\sqrt{1+d_3 R^2}} ]
\label{d9}
\end{equation}
It also can be seen that Equs. (\ref{eq11} - \ref{eq12}) are satisfied identically.  we have from equ. (\ref{eq16}), the acceleration
\begin{equation}
a = \frac{d_1 R}{\int \frac{d_1}{R} dr +d_2}
\label{d10}
\end{equation}
Thus the solutions presented here have non-zero acceleration and dissipation which reveals the dynamical cavity model of star.
%%%%%%%%%%%%%%%%%%%%%%%%%%%%%%%%%%%%%%%%%%%%%%%%%%%%%%%%%%%%%%%%%%%%%

%%%%%%%%%%%%%%%%%%%%%%%%%%%%%%%%%%%%%%%%%%%%%%%%%%%%%%%%%%%%%%%%%%%%%%%%%%%%%
%%%%%%%%%%%%%%%%%%%%%%%%%%%%%%%%%%%%%%%%%%%%%%%%%%%%%%%%%%%%%%%%%%%%%%%%%%%%%%%%%%%%%%%%%%%%
\section{Self-similar regime}
The existence of a homothetic- killing vectors described the self similar space-time. The self similar solutions of the field equations in general relativity was studied in extensive details ( see \cite{ct71}, \cite{jd92} and references their in ).  Any spherically-symmetric space-time is self similar if it contain a radial- coordinate $r$ and an orthogonal time-coordinate $t$ so that for the metric-coefficient $g_{tt}$ and $g_{rr}$ satisfied
\begin{equation}
g_{tt}(kt, kr) = g_{tt} (t,r)
\end{equation}
\begin{equation}
g_{rr}(kt, kr) = g_{rr} (t,r)
\end{equation}
for all constant $k> 0$ . In  the self similar regime, the field equations, a set of partial-differential equation transform into ordinary-differential equation.  Self similarities are the strong constraint of geometry, which have been effectively expressed in various cosmological/physical scenarios (\cite{ct71} - \cite{wg06}). In this section it has studied that the appearance of a homothetic-killing vector field for a spherically-symmetric space-time entails the detachability of the space time metric components in terms of the comoving coordinates and that the line element can be expressed in a simplified unique form. A vector field $\xi^i$ is said to be homothetic if it follows

\begin{equation}
\mathcal{L}_\xi g_{ij} = 2g_{ij}
\label{ss0}
\end{equation}
Consider that a spherically-symmetric space-time consist a homothetic-killing vector of the form
\begin{equation}
\xi^i = (0, \alpha(t,r), 0, 0)
\label{ss1}
\end{equation}
Usually, a homothetic-Killing vector is expressible in the form
\begin{equation}
\xi^i = (\overline{t}, \overline{r}, 0, 0)
\label{ss2}
\end{equation}
However, any vector of the form (\ref{ss1}) can be transformed into the form (\ref{ss2}) via a coordinate
transformation(\cite{cc99} - \cite{wg06})
\begin{equation}
\overline{t} = l(t) e^{\int \alpha^{-1} dr},\quad~ \overline{r} = k(t) e^{\int \alpha^{-1} dr}
\label{ss3}
\end{equation}
If the  line element (\ref{eq6})  admits a homothetic killing vector of the form (\ref{ss1}), then equ. (\ref{ss0}) yields
\begin{equation}
\frac{A'}{A} = 4 \alpha^{-1}
\end{equation}
\begin{equation}
\dot{\alpha}(t,r) = 0
\end{equation}
\begin{equation}
\alpha' = 2(2+\alpha \frac{R'}{R})
\end{equation}
\begin{equation}
\frac{R'}{R} = 4 \alpha^{-1}
\end{equation}
Solving the above systems of equations, one obtain
\begin{equation}
\alpha = 12r
\end{equation}
\begin{equation}
A(t,r)= k(t) r^{\frac{1}{3}}
\label{ss4}
\end{equation}
\begin{equation}
R(t,r)= l(t) r^{\frac{1}{3}}
\label{ss5}
\end{equation}
where $k(t)$ and $l(t)$ are arbitrary  constants. In view of Equ. (\ref{ss4}) and (\ref{ss5}), the metric (\ref{eq6} )reduces to the form
\begin{equation}
ds_{-}^2 = -k^2 r^{\frac{2}{3}} dt^2 + l^{-4} r^{\frac{-4}{3}} dr^2 + l^2 r^{\frac{2}{3}} (d\theta^2 +\sin\theta^2 d\phi^2)
\label{ss6}
\end{equation}
Thus, by virtue of this it can be obtain from equation (\ref{eq7}) - (\ref{eq10})
\begin{equation}
8 \pi \rho  = \frac{1}{3 r^{\frac{2}{3}} k^3 l^2} [ (-l^6+6) k^3 -9 l \dot{k} \dot{l}+9k(l \ddot{l} -2 \dot{l}^2)  ]
\label{ss7}
\end{equation}
\begin{equation}
8 \pi p  = \frac{1}{9 r^{\frac{2}{3}} k^3 l^2}[ k^3 l^6-9 l\dot{l}\dot{k} +9 k(l \ddot{l}-4\dot{l}^2) ]
\label{ss8}
\end{equation}
\begin{equation}
8 \pi q  = \frac{1}{9 r^{\frac{2}{3}} k^3 l^2}[ k^3(9-2 l^6)+12 k^2 l^3 \dot{l} -27 l\dot{l} \dot{k}-27 k (\dot{l}^2- l\ddot{l}) ]
\label{ss9}
\end{equation}
\begin{equation}
8 \pi \epsilon  = \frac{1}{9 r^{\frac{2}{3}} k^3 l^2}[k^3 (2l^6-9) + 27 l\dot{l} \dot{k} + 27k(\dot{l}^2-l\ddot{l}) ]
\label{ss10}
\end{equation}
Also, the equations (\ref{eq13}) and (\ref{eq17}) yield
\begin{equation}
m(t,r) = \frac{1}{2} r^{\frac{1}{3}} l(1-\frac{l^6}{9}+\frac{\dot{l}^2}{k^2})
\label{ss11}
\end{equation}
\begin{equation}
\sigma = -3\frac{\dot{l}}{r^{\frac{1}{3}} k l}
\label{ss12}
\end{equation}
It follows from equ. (\ref{eq16}), the acceleration
\begin{equation}
a = \frac{1}{3} l^2 r^{\frac{-1}{3}}
\label{ss13}
\end{equation}
Now, apply the junction condition (\ref{eq27}) over hypersurface $\Sigma^e$
\begin{equation}
k(t)= -\frac{(M-4) \dot{M}}{3^{2/3} \sqrt{\frac{(4-3 M)^2}{M-2}} (M-2)^{7/6} M^{2/3}}
\label{ss14}
\end{equation}
\begin{equation}
l(t) = \frac{\sqrt[3]{3} \sqrt[3]{M(t)}}{\sqrt{2} \sqrt[6]{M(t)-2}}
\label{ss15}
\end{equation}
and thus the arbitrary functions $l(t)$ and $k(t)$ are completely determined. Also, from equations (\ref{eq28}), (\ref{ss8}), (\ref{ss9}) and (\ref{ss11}) it can observed that the model (\ref{ss6}) does not satisfies the Darmois junction condition over cavity hypersurface $\Sigma^i$. Therefore, it might be eprtinent to relax jucntion condition and it exhibits the presence of thin shell which allows for the discontinuities of mass function across $\Sigma^i$.
%%%%%%%%%%%%%%%%%%%%%%%%%%%%%%%%%%%%%%%%%%%%%%%%%%%%%%%%%%%%%%%%%%%%%%%%%%%%%%%%%%%%%%%
\section{Discussion and Concluding remarks}
In last decade the evolution of expansion free dynamical stars have taken considerable interest among relativists and has been applied to illustrate the physical and geometrical properties of radiating star (\cite{hds08}-\cite{phosc11}, \cite{hgs12}- \cite{ks2018a} and references their in). It can be emphasized that  very limited dynamical models are investigated for dissipative cavity evolution.  The existing work is in itself very remarkable to describe the cavity model of self-gravitating spherically symmetric dissipative perfect fluids. For obtaining solution, some alternatives are presented namely, quasi-static diffusion approximation  and self–similar regime to elucidate the modelling of cavity evolution.

\par
In this way, the first solution is investigated with diffusion approximation with quasi-static regime and it has been shown  that the energy density is homogeneous, $\rho = \rho(t)$ and   $\Lambda-$ dominated (with negative energy density which may be violated  baryonic equation of state,\cite{y17}).  As an important special case, self-similarity (homothetic) is also introduced here to discuss the solution of field equations. It is interesting to note that the model is explicitly described in terms of $r$ and arbitrary function $k(t), l(t)$. Since it is  seeing here describing the localized objects without the unusual topology of a spherical fluid without centre, $r \neq 0$, the centre is surrounded by a compact spherical surface of another space time suitably matched to the rest of the fluid\cite{hds08}, \cite{hpo10}. Therefore no real singularity occurs in this model at all ($r\neq 0$). Possibly the solutions presented here  could be applied for the localized system of supernova explosion. Forthcoming researches of such models with numerical solutions of pertinent equation would produce more appropriate applications in astrophysics.

%%%%%%%%%%%%%%%%%%%%%%%%%%%%%%%%%%%%%%%%%%%%%%%%%%%%%%%%%%%%%%%%%%%%%%%%%%%%%%%%%%%%%%%%%%%%%%%%%%%%%

%%%%%%%%%%%%%%%%%%%%%%%%%%%%%%%%%%%%%%%%%%%%%%%%%%%%%%%%%%%%%%%%%%%%%%%%%%%%%%%%%%%%%%%%%%
\end{document}